\documentclass[12pt,a4paper,final]{iopart}
\pdfoutput=1

\usepackage{iopams}
\usepackage{amssymb}
\usepackage{graphicx,caption,cite}

\usepackage[colorlinks,allcolors=blue]{hyperref}

\usepackage{tikz}
\usepackage{lineno}

\usepackage{wasysym}
\usepackage{eucal}
\usepackage{amsfonts}

\begin{document}


  \title{From continuous-time random walks to controlled-diffusion reaction}
\author{Maike A. F. dos Santos}

\address{Instituto de F\'isica, Universidade Federal do Rio Grande do Sul, Caixa Postal 15051, CEP 91501-970, Porto Alegre, RS, Brazil}
\ead{santosmakeaf@gmail.com}
\vspace{10pt}
\begin{indented}
\item[] \today
\end{indented}

\begin{abstract}

Daily, are reported systems in nature that present anomalous diffusion phenomena due to irregularities of medium, traps or reactions process. In this scenario, the diffusion with traps or localised--reactions emerge through various investigations that include numerical, analytical and experimental techniques. In this work, we construct a model which involves a coupling of two diffusion equations to approach the random walkers in a medium with  localised reaction point (or controlled diffusion). We present the exact analytical solutions to the model. In the following, we obtain the survival probability and mean square displacement. Moreover, we extend the model to include memory effects in reaction points. Thereby, we found a simple relation that connects the power-law memory kernels with anomalous diffusion phenomena, i.e. $\langle (x-\langle x \rangle )^2 \rangle \propto t^{\mu}$. The investigations presented in this work uses recent mathematical techniques to introduces a form to represent the coupled random walks in context of reaction-diffusion problem to localised reaction.

${ }$

\end{abstract}

\maketitle


\section{\label{sec1}Introduction}

Nowadays, the investigation of the diffusion process is one of the most applicable concepts of statistical physics which includes fields like Biology, Medicine and Chemistry \cite{capasso2005introduction,polyanin2016handbook,britton1986reaction}. The modern scenario of microscopic diffusion began with investigations of botanic Robert Brown about grains of pollen of the plant \textit{Clarkia pulchella} \cite{brown1828xxvii}.  Brown experiment's revealed the irregular movement of \textit{amyloplasts} and \textit{spherosomes} (particles contained in pollen grains) immerse in water. This phenomenon known as Brownian motion (BM) was described in 1905 on Einstein's work \cite{einstein1905molekularkinetischen}. Thereby, other formulations were made by Sutherland, Smoluchowski, and Langevin \cite{sutherland1905lxxv,von1906kinetischen,langevin1908theorie}. These works characterized the BM by a linear evolution of mean square displacement (MSD), i.e. $\langle (\Delta x)^2 \rangle \sim t$. Between 1908 and 1914, Perrin and Nordlund published experimental results that proved statistical models of Brownian motion \cite{perrin1908agitation,perrin1909mouvement,nordlund1914new}. In this context, the diffusion equation introduced by Einstein plays an important role in statistical physics of the microscopic objects. Therefore, the Brownian motion presented on pollen grains is a simple consequence of a complex movement of multiplies interactions of atoms and molecules. These ideas converged to consolidate the existence of atoms and molecules.

Inspired by these ideas, 
in 1914, Herzog e Polotzky \cite{herzog1914diffusion} realised experiments to analyze the Fick’s laws. Nevertheless, only in 1935, Freundlich and Kr{\"u}ger presented that these experiments culminate in deviations of the usual diffusion process \cite{freundlich1935anomalous}. These reports implied the beginning of studies  on anomalous  diffusion, i.e., $\langle (\Delta x)^2 \rangle  \nsim t$.
 In 1926, Richardson investigated the relative diffusion of two tracer particles in turbulent flows to systems ranging from capillary tubes to cyclones \cite{richardson1926atmospheric}. In 1975, Scher and Montroll \cite{scher1975anomalous} related the anomalous diffusion in the dispersive transport of charge carrier motion in amorphous semiconductors, as approached by Weiss and Montroll an investigation of continuous time random walk (CTRW) \cite{montroll1965random}.
 The Richardson and Montroll work culminated in a  classification of anomalous diffusion process by power-law behaviour of the MSD, as follows 
\begin{eqnarray}
\langle (\Delta x)^2 \rangle = 2 \mathcal{K}_{\mu} t^{\mu},
\label{anomalous}
\end{eqnarray}
in which $\mathcal{K}_{\mu}$ is general diffusion coefficient with fractional dimension $[\mathcal{K}_{\mu}]=\textnormal{cm}^2/ \textnormal{[t]}^{\mu}$. Actually, the relation (\ref{anomalous}) is associated with multiples diffusive behaviour, classified as: $0<\mu<1$ the system is sub-diffusive, $\mu=1$ usual diffusion, to $1<\mu<2$ occurs the super-diffusion. In particular cases, to $\mu=2$ the diffusion is ballistic and for $2<\mu$ occurs the hyper diffusive process. The anomalous diffusion has been  reported oftentimes in statistical mechanics out of equilibrium that usually implies non-Gaussian distributions \cite{barkai,chechkin2017brownian,sposini2018random,metzler2000random,slkezak2018superstatistical,buonocore2018tomographic}.

In this decade, the mechanisms that imply the anomalous diffusion have received more attention \cite{hristov2017derivatives,magdziarz2011anomalous,yang2017new,zhang2017nontrivial}. Curiously, there are several biological systems in which the collective movement needs some type of special dynamical approach, which can include simulation, experiment and analytical models. An example  is in recent work of Pan Tan \textit{et al.}, \cite{PhysRevLett.120.248101} that shows a series of experimental results which reveal a  subdiffusive behaviour to water on hydration surface of proteins. The authors suggest that the water near of proteins behaves as if the protein were cages that eventually trap the particles, as if the surface were a fractal system. In this sense, Ralf Metzler discusses how different techniques can reveal a different \textit{dance of water} (anomalous diffusion) around proteins  \cite{metzler2018dance}. Hence, investigate the mechanisms which lead to anomalous behaviour is a central theme in statistical mechanics.  In the article \cite{barkaiIntro}, the authors present  the main formalisms to approach  anomalous  diffusion.

\begin{figure}[h]
\centering
\includegraphics[scale=1]{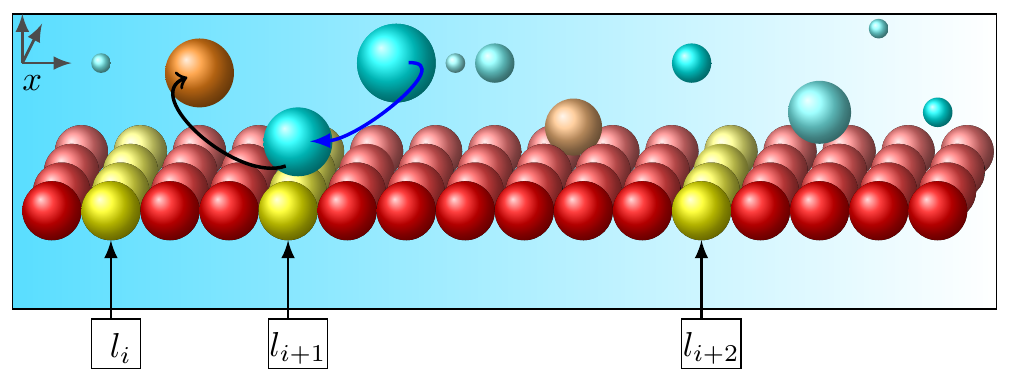}
\caption{\small{The blue particles (type 1) and orange particles (type 2) represent substances near of the surface \textcolor{red}{\CIRCLE}. The surface have specific places (lines) denoted by positions $x=l_i$ (\textcolor{yellow}{\CIRCLE}), that represent the places in which localised reaction occurs a irreversible (or reversible) reaction between walkers 1 ($\textnormal{\textcolor{cyan}{\CIRCLE}})$ and 2 ($\textnormal{\textcolor{orange}{\CIRCLE}}$). 
}}
\label{particulas2}
\end{figure}

As it was pointed by many authors, the diffusion in irregular or fractal environments tends to present the anomalous behaviour to diffusion  \cite{metzler2000random,tarasov2005fractional,tarasov2005fractional2,buonocore2018tomographic}. On the other hand, there are several  systems in which the diffusion of atoms or molecules occurs in a medium with trap--points (or obstacles)  \cite{sung2002fractional,krall1998internal}. This problem can be more difficult if we consider that occurs a localised reaction between two types of particles. Approach this kind of system is relevant to the investigation of irregularities influence of medium in the diffusive process. In this sense, some models to localised reaction (or controlled reaction) were proposed \cite{szabo1984localized}, but no exact analytic solutions. Recently, some works have approached the diffusion in a medium with a single point (or the reaction point), in contexts as search strategies for single and multiple targets \cite{whitehouse2013effect,palyulin2017comparison,palyulin2014levy} and minimal model \cite{flekkoy2017minimal}. In contrast,  studies on the Refs.  \cite{sandev2017heterogeneous,sandev2017anomalous} introduced a way to approach the diffusion in systems with geometric restrictions associated with fractal sets. In this scenario, we propose a diffusion--reaction model to approach the complex controlled-reaction. The model investigated in this work may introduce new forms to investigate the precipitation phenomena \cite{fahey1989point,sahai2000x,alam2015structural,TU2018133} and traps-system \cite{szabo1984localized,zhou2018molecular,freidlin2017behavior,moss2017comparing,lindsay2017optimization} with exact solutions. 
To do this, we consider a localised reaction point (idea in Fig.  \ref{particulas2}), in which the reaction-point are places in that particles of type 1 change to type 2 or vice versa.
The model implies a series of quantities that can be determined thought analytical calculus of probability distribution, as MSD and survival probability.

The paper is outlined as follows: in section  \ref{sec2}, we use the CTRW theory to introduce our model that consists of a system of the diffusion equations which are coupled by localised reaction term. In section \ref{sec3} we present the exact solution for the problem in markovian case. In the following, we present a series of behaviour to probability density function associated with particles spreading on the system. In section \ref{sec31}, we obtain exact expressions for the survivals probability and MSD. As consequence, we show that the anomalous diffusion emerges due presence of reaction point, i.e. $\langle x^2 \rangle \propto t^{\mu}$. Moreover, in section \ref{sec4}, we present the model with memory effects. Finally, in section \ref{conclusion}, we present the conclusions and futures possibilities to be investigated in this scenario.

\section{\label{sec2} The coupled random walkers}

The diffusion equation has one of the most robust mathematical structures investigated in nature. In this sense, we can write a equation to random walkers which include the reversible reaction process between walkers of type $1$ or $2$, according to the rule $\textnormal{1} \leftrightarrows \textnormal{2}$. These systems are typically described by coupling of the diffusion equations, which is an approach  very well establishes on chemical--physic problems \cite{polyanin2016handbook}. Mentioning some examples, we have coupled CTRW \cite{langlands2008anomalous}, chemical systems \cite{iyiola2017analytical,kopelman1988fractal}, non-linear dynamic \cite{jiwari2017numerical,polyanin2016handbook}. We will propose a model describe two walkers as illustrating on Fig. (\ref{particulas2}), represented by irreversible reaction process, 
$\mathcal{W}_1(x,t) \leftrightarrows \mathcal{W}_2(x,t)$, to particles of type 1 and 2 respectively, in one-dimensional space ($x$--axis).

The diffusion equations to walkers $1$ and $2$ can be obtained by means of the integral equation for the CTRW \cite{metzler2000random}, obtained by means of the Fourier transform, following procedure. It is considered the average waiting time,
\begin{eqnarray}
\tau = \int_0^\infty t \psi(t) dt ,
\end{eqnarray}
and the jump length variance $
\sigma^2 = \int_{-\infty}^{\infty} x^2 \lambda(x) dx $ to stochastic steps. 
By means of such averages, we can characterise different types of CTRW considering the finite or divergent nature of these quantities. Our propose  considers two coupled CTRW as follows
\begin{eqnarray}
\eta_1(x, t) &=& \int_{-\infty}^\infty  \; \int_0^t dx' dt' \eta_1(x', t') \Psi_1(x-x', t-t') \nonumber \\  &+& \mathcal{W}_1(x,0)   \delta(t) + \mathcal{R}_{1,2},
\label{1} \\
\eta_2(x, t) &=& \int_{-\infty}^\infty  \; \int_0^t dx' dt' \eta_2(x', t') \Psi_2(x-x', t-t')  \nonumber \\  &+& \mathcal{W}_2(x,0)   \delta(t) + \mathcal{R}_{2,1},
\label{2}
\end{eqnarray}
   $\eta_i(x, t)$ is the probability per unit of displacement and time of a random walker who has left the $x$ in the time $t$, to the position $x'$ in time $t'$, being the last term (product of two deltas) the initial condition of the walker.

Therefore, the probability density function $\mathcal{W}_i(x, t)$ of the walker to be found in $x$ in the time $t$ is given by
\begin{eqnarray}
\mathcal{W}_i(x, t) = \int_0^t \eta_i(x, t') \; \Phi_i(t-t') dt',
\label{11,2,4}
\end{eqnarray}
in which $i \in \{1,2 \}$ and 
\begin{eqnarray}
\Phi_i(t) = 1- \int_0^t \psi_i(t') \; dt'
\label{11,2,5}
\end{eqnarray}
 is the probability of the walker not jumping during the time interval $(0,t)$, that is, to remain in the initial position. Applying the Laplace transform  ($\mathcal{L}\{ f(t) \}=\int_0^{+\infty}e^{-st}f(t) dt=\tilde{f}(s)$) in equations (\ref{11,2,4}),  (\ref{11,2,5}) and using the convolution theorem, we have
\begin{eqnarray}
\widetilde{\mathcal{W}}_i(x, s) = \frac{1}{s} \widetilde{\eta}_i(x, s) [1-\psi_i(s)] .
\label{11,2,7}
\end{eqnarray}
To determinate $\widetilde{\eta}(x,s)$, we must return to (\ref{1}, \ref{2}) and apply the Laplace transform on the temporal and Fourier variables on the spatial variable ($\mathcal{F}\{ g(x) \}=\int_{-\infty}^{+\infty}e^{-i k x}g(x) dx=g(k)$). Making use of integral transformations, we have
\begin{eqnarray}
\widetilde{\eta}_i(k, s) [1-\Psi_i(k, s)]=\mathcal{W}_i(k,0)+\widetilde{\mathcal{R}}_{i,j}(k,s),
\end{eqnarray}
in which $i \in \{1,2\}$, $j \in \{1,2\}$, $i \neq j$.

Using the previous result and considering a generic initial condition $\mathcal{W}_0 (x)$, we have
\begin{eqnarray}
\widetilde{\mathcal{W}}_i (k, s) = \frac{1-\psi_i(s)}{s}\frac{\mathcal{W}_i(k,0)+\widetilde{\mathcal{R}}_{i,j}(k,s)}{1-\Psi_i(k, s)}.
\label{distributionctrw}
\end{eqnarray}
\noindent This equation can be applied to systems that have the jump length coupled to the waiting time.  Here we can introduce  a waiting time distribution which describes a more general class of  Random walkers
\begin{eqnarray}
\Psi_i(k,s)= \psi_i(s)\lambda_i(k),
\label{w1}
\end{eqnarray}
in that 
\begin{eqnarray}
\psi_i(s)= \frac{\mathcal{M}_i(s)}{\mathcal{M}_i(s)+\tau_i s}
\end{eqnarray}
and $\lambda_i(k)=\exp[-\frac{k^2}{2\sigma_i^2}]$, to asymptotic limit we have  $\lambda_i(k) \sim 1 - \frac{\sigma_i^2}{2}k^2$. That in Eq. (\ref{distributionctrw}) we obtain
\begin{eqnarray}
\widetilde{\mathcal{W}}_i (k, s) = \frac{\mathcal{W}_i(k,0)+\widetilde{\mathcal{R}}_{i,j}(k,s)}{ s +\mathcal{M}_i(s)\frac{\sigma_i^2}{2 \tau_i}k^2}.
\label{distributionctrw}
\end{eqnarray}
assuming the quantities 
\begin{eqnarray}
\widetilde{\mathcal{K}}_i(s)=\mathcal{M}_i(s)\frac{\sigma_i^2}{2 \tau_i }.
\end{eqnarray}
Thereby, we  write the dynamical equations to systematize as follow
\begin{eqnarray}
\frac{\partial \ }{\partial t}\mathcal{W}_1(x,t) = \int_0^t dt' \mathcal{K}_1(t-t')\frac{\partial^2}{\partial x^2} \mathcal{W}_1(x,t') + \mathcal{R}_{1,2}(x,t),
\label{sys1f}
\\
\frac{\partial \ }{\partial t} \mathcal{W}_2(x,t)= \int_0^t dt'  \mathcal{K}_2(t-t')\frac{\partial^2}{\partial x^2}  \mathcal{W}_2(x,t') + \mathcal{R}_{2,1}(x,t),
\label{sys2f}
\end{eqnarray}
this set of equations represent a most sophisticate interaction form between two type of walker. The equations (\ref{sys1f}) and (\ref{sys2f}) are very general, to $\mathcal{K}_i(t)=\delta(t)$ they  cover a series of well know situations \cite{kotomin1996modern};  to $\mathcal{R}_{i,j}(x,t)=-k_i f(\mathcal{W}_i(x,t))$ we obtain two uncoupled reaction-diffusion equations  with memory \cite{Yadav}. But here, we will put the problem in context of controlled reaction, that consist in investigate the diffusion of walkers where the coupled-term occurs in a single point, as considered in general theory of controlled-diffusion reaction  \cite{wilemski1973general}. 

\section{\label{sec3} Controlled-diffusion  and localised reaction }

In 1984 Szabo \textit{el al.} proposes a model to diffusion with localised traps for one specie of particle $\mathcal{R}= -\delta(x) k\rho(x,t)$, see Refs. \cite{szabo1984localized}  and \cite{sung2002fractional}. Szabo's model is bound to  Wilemski and Fixman's ideas, presented in the general theory of diffusion--controlled reactions \cite{wilemski1973general}. 

The model proposed by us in previous section consider a coupled diffusion equation. We start our discuss considering the non-memory case, i.e. $\mathcal{K}_i(t) \propto \delta(t)$, that may be resumed by expressions
\begin{eqnarray}
\frac{\partial \ }{\partial t}\mathcal{W}_1(x,t) = \mathcal{K}_1\frac{\partial^2}{\partial x^2} \mathcal{W}_1(x,t) + \mathcal{R}_{1,2}(x,t),
\label{sys1}
\\
\frac{\partial \ }{\partial t} \mathcal{W}_2(x,t)= \mathcal{K}_2\frac{\partial^2}{\partial x^2}  \mathcal{W}_2(x,t) + \mathcal{R}_{2,1}(x,t),
\label{sys2}
\end{eqnarray}
in that $\mathcal{K}_1$ and $\mathcal{K}_2$ are diffusion coefficients and $\mathcal{R}_{i,j}$ represent the reaction term. The quantity $\mathcal{R}_{i,j}$ represents the rate in which one type of walkers are converted into another and vice-versa. In this sense, the linear choice to reaction term, $\mathcal{R}_{1,2} =-k_1\mathcal{W}_1 + k_2 \mathcal{W}_2$, 
defines a uniform reaction process in space and was extensively investigated in literature \cite{henry2000fractional,henry2006anomalous,dos2017anomalous,dos2017nonlinear,kindermann2015data,ciocanel2017analysis}. Here, we propose that a reaction process between two species occurs in a specific point which  defines a controlled-reaction for two species.  Thereby, we begin defining the term of transition proposed by us, as follows
 \begin{eqnarray}
 \mathcal{R}_{i,j} = \delta(x) \left\lbrace -k_i\mathcal{W}_i(x,t)+k_j \mathcal{W}_j(x,t) \right\rbrace,
 \end{eqnarray}
 in which $i \in \{1,2\}$, $j \in \{1,2\}$, $i \neq j$.
 Considering the initial conditions
\begin{eqnarray}
\mathcal{W}_1(x,0)=\delta(x-l), \qquad \textnormal{and} \qquad \mathcal{W}_2(x,0)= 0,
\label{initialconditions}
\end{eqnarray}
and boundaries conditions
\begin{eqnarray}
\mathcal{W}_1(\pm \infty , 0)=0, \qquad \textnormal{and} \qquad \mathcal{W}_2(\pm \infty , 0)=0.
\label{boundariesconditions}
\end{eqnarray}
We can rewritten the Eqs. (\ref{sys1} and \ref{sys2}) in Fourier-Laplace space, one obtains
\begin{eqnarray}
s \widetilde{\mathcal{W}}_1(k,s)=  -k^2\mathcal{K}_1 \widetilde{\mathcal{W}}_1(k,s)+\delta(x-l)+\widetilde{\mathcal{R}}_{1,2}(k,s),
\label{diffusion1}
\end{eqnarray}
and
\begin{eqnarray}
s \widetilde{\mathcal{W}}_2(k,s) = -k^2\mathcal{K}_2 \widetilde{\mathcal{W}}_2(k,s) 
 + \widetilde{\mathcal{R}}_{2,1}(k,s),
 \label{diffusion2}
\end{eqnarray}
with $\widetilde{\mathcal{R}}_{1,2}= \delta(x)  \left\lbrace -k_1 \widetilde{\mathcal{W}}_1(x,s)+ k_2\widetilde{\mathcal{W}}_2(x,s) \right\rbrace $ and $\widetilde{\mathcal{R}}_{1,2}=-\widetilde{\mathcal{R}}_{2,1}$. We found the solution in Fourier-Laplace space to diffusion equations, we obtain
\begin{eqnarray}
 \widetilde{\mathcal{W}}_1(k,s)&=& \frac{1}{s+k^2 \mathcal{K}_1}(e^{- i k l}-k_1 \widetilde{\mathcal{W}}_1(0,s)+k_2 \widetilde{\mathcal{W}}_2(0,s)),  \label{ansatz}
 \\
  \widetilde{\mathcal{W}}_2(k,s) &=&\frac{1}{s+k^2 \mathcal{K}_2}(-k_2 \widetilde{\mathcal{W}}_2(0,s)+k_1 \widetilde{\mathcal{W}}_1(0,s)).
  \label{ansatz2}
\end{eqnarray}
In this part we need found the  $\widetilde{\mathcal{W}}_1(0,s)$ and $\widetilde{\mathcal{W}}_1(0,s)$ functions. 
Realising the inverse Fourier transform in Eqs. (\ref{ansatz} and \ref{ansatz2}) and assuming $x=0$, we obtain
\begin{eqnarray}
 \widetilde{\mathcal{W}}_1(0,s)&=& \frac{1}{2\sqrt{s \mathcal{K}_1}} (e^{-\sqrt{\frac{s}{\mathcal{K}_1}}|l|}-k_1 \widetilde{\mathcal{W}}_1(0,s)+k_2 \widetilde{\mathcal{W}}_2(0,s)),  \label{ansatzm}
 \\
  \widetilde{\mathcal{W}}_2(0,s) &=& \frac{1}{2\sqrt{s \mathcal{K}_2}} (-k_2 \widetilde{\mathcal{W}}_2(0,s)+k_1 \widetilde{\mathcal{W}}_1(0,s)).
  \label{ansatz2m}
\end{eqnarray}
rewritten the Eq. (\ref{ansatz2m})  
\begin{eqnarray}
\widetilde{\mathcal{W}}_2(0,s)= \frac{1}{2\sqrt{s\mathcal{K}_2}+k_2} k_1\widetilde{\mathcal{W}}_1(0,s),
\label{w2teste}
\end{eqnarray}
replacing the Eq. (\ref{w2teste}) in Eq. (\ref{ansatzm}) obtains
\begin{eqnarray}
\widetilde{\mathcal{W}}_1(0,s) &=&  \frac{e^{-\sqrt{\frac{s}{\mathcal{K}_1}}|l|}}{2\sqrt{s\mathcal{K}_1}+ k_1 - \frac{k_2k_1}{2\sqrt{s\mathcal{K}_2}+k_2}} \nonumber \\
&=& e^{-\sqrt{\frac{s}{\mathcal{K}_1}}|l|}\frac{2\sqrt{s\mathcal{K}_2}+k_2}{(2\sqrt{s\mathcal{K}_1}+ k_1)(2\sqrt{s\mathcal{K}_2}+k_2) - k_2k_1}  \nonumber \\
&=& e^{-\sqrt{\frac{s}{\mathcal{K}_1}}|l|}\frac{2\sqrt{s\mathcal{K}_2}+k_2}{4 s \sqrt{\mathcal{K}_1\mathcal{K}_2}+2\sqrt{s}(\sqrt{\mathcal{K}_1}k_2+\sqrt{\mathcal{K}_2}k_1)},
\label{incognita1}
\end{eqnarray}
thereby
\begin{eqnarray}
\widetilde{\mathcal{W}}_2(0,s)= e^{-\sqrt{\frac{s}{\mathcal{K}_1}}|l|}\frac{k_1}{4 s \sqrt{\mathcal{K}_1\mathcal{K}_2}+2\sqrt{s}(\sqrt{\mathcal{K}_1}k_2+\sqrt{\mathcal{K}_2}k_1)}.
\label{incognita2}
\end{eqnarray}
Replacing the Eqs. (\ref{incognita1} and \ref{incognita2}) in Eqs. (\ref{ansatz} and \ref{ansatz2}). We obtain the exact solutions in Fourier-Laplace space
\begin{eqnarray}
 \widetilde{\mathcal{W}}_1(k,s)&=& \frac{1}{s+k^2 \mathcal{K}_1} \left(e^{-i l k}-e^{-\sqrt{\frac{s}{\mathcal{K}_1}}|l|} \Upsilon(s) \right) \label{w11} \\
\widetilde{\mathcal{W}}_2(k,s)&=& \frac{1}{s+k^2 \mathcal{K}_1}\left( e^{-\sqrt{\frac{s}{\mathcal{K}_1}}|l|} \Upsilon(s) \right), \label{w22}   
\end{eqnarray} 
  with 
\begin{eqnarray}
\widetilde{\Upsilon}(s) = \frac{1}{2\sqrt{\mathcal{K}_1}}\frac{k_1}{\sqrt{s}+\gamma},
\end{eqnarray}
in which the constant quantity is $\gamma= 2^{-1}(\frac{k_2}{\sqrt{\mathcal{K}_2}}+\frac{k_1}{\sqrt{\mathcal{K}_1}})$.

\begin{figure}[h!]
\centering
\includegraphics[scale=0.8]{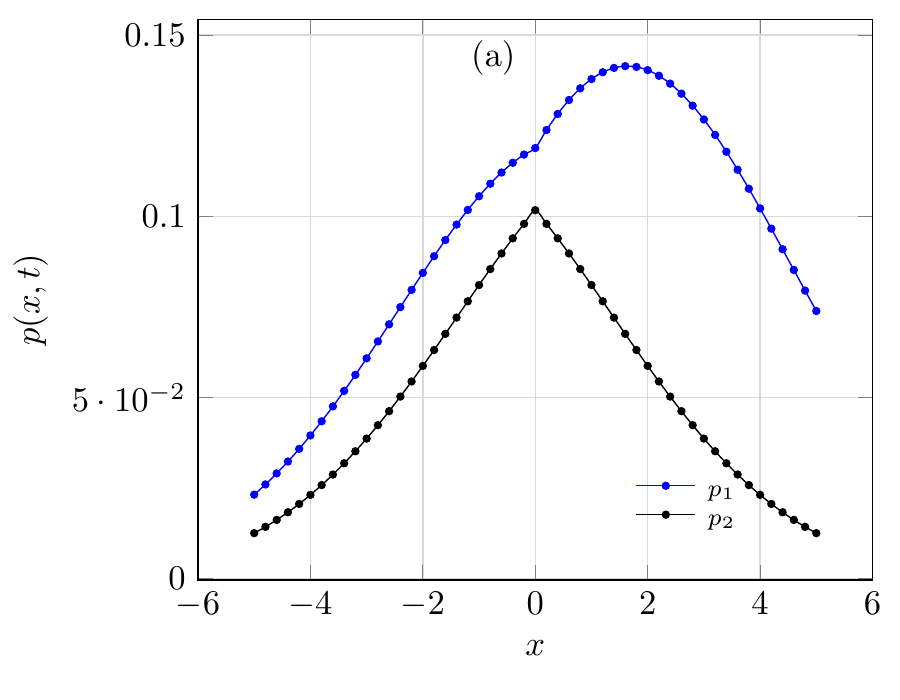}\\
\includegraphics[scale=0.8]{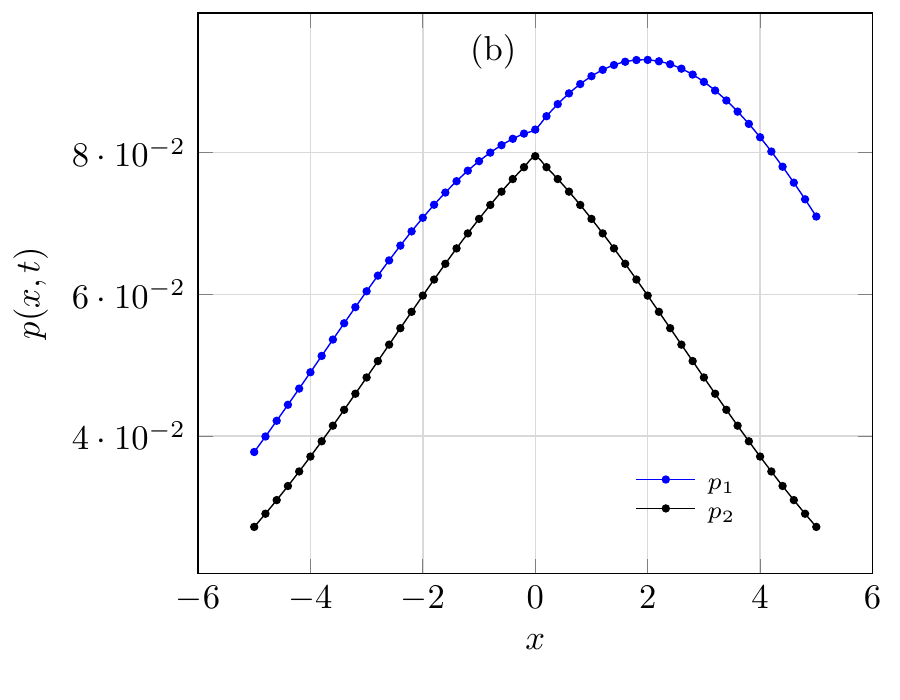}\\
\includegraphics[scale=0.8]{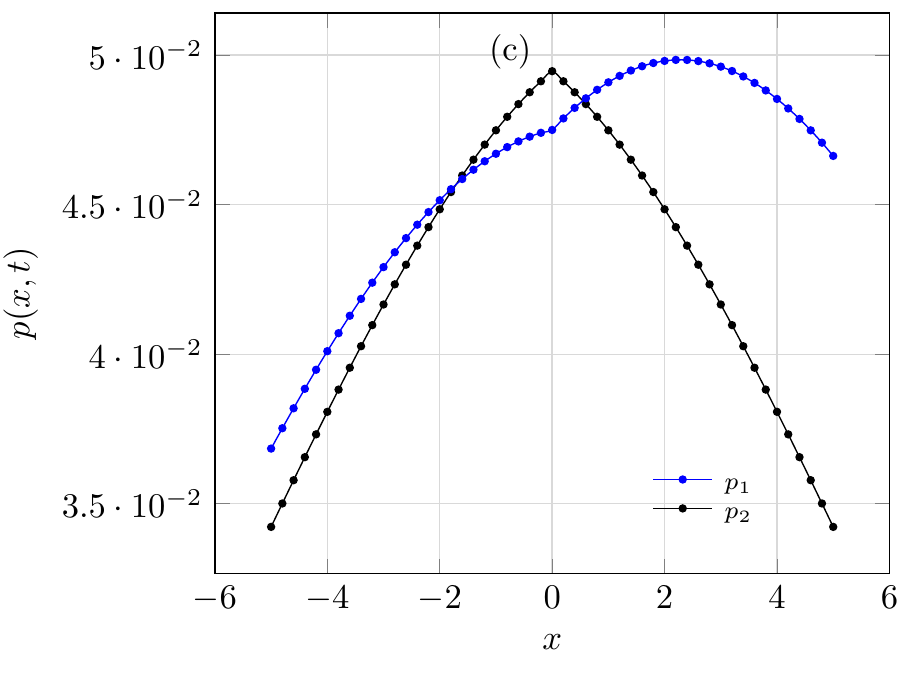}
\caption{\small{The figures present the probability density distribution to both walkers, for different values of time, $t=5$ in fig. (a), $t=10$ in fig. (b) and $t=30$ in fig. (c). Considering $l=1$,  $k_1=k_2=1$ and $\mathcal{K}_1=\mathcal{K}_2=1$.
}}
\label{fig2}
\end{figure}

In view of the above considerations, we must find the  $\Upsilon(t)$-function.
Firstly, performing  the inverse Fourier transform have the solution formed of Eqs. (\ref{w11}) and (\ref{w22}) in Laplace space.
 We obtain
\begin{eqnarray}
\widetilde{\mathcal{W}}_1(x,s) &=&  \frac{1}{2\sqrt{s\mathcal{K}_1}} \exp \left[ -\sqrt{\frac{s}{\mathcal{K}_1}}|x-l| \right] \nonumber \\ 
&-& \widetilde{\Upsilon}(s) \frac{1}{2\sqrt{s\mathcal{K}_1}} \exp \left[ -\sqrt{\frac{s}{\mathcal{K}_1}}(|x|+|l|) \right],
\label{solution11}\\
\widetilde{\mathcal{W}}_2(x,s) &=&  \frac{\widetilde{\Upsilon}(s)}{2\sqrt{s\mathcal{K}_2}}  \exp \left[ -\sqrt{\frac{s}{\mathcal{K}_2}}|x| -\sqrt{\frac{s}{\mathcal{K}_1}}|l| \right] \label{solution12}. 
\end{eqnarray}
To $k_1=0$ we recover the Gaussian solution to particles of type 1. 
 Considering the general situation $k_1\neq 0$ and $k_2\neq 0$ the inversion in Laplace transform of Eq. (\ref{solution11}) implies solution to systems, written as
\begin{eqnarray}
 \mathcal{W}_1(x,t) = \frac{\exp \left[-\frac{(x-l)^2}{4 \mathcal{K}_1 t} \right]}{2\sqrt{\pi t \mathcal{K}_1}} - \mathcal{L}^{-1} \bigg \{ \widetilde{\Upsilon}(s) \frac{\exp \left[ -\sqrt{\frac{s}{\mathcal{K}_1}}(|x|+|l|) \right]}{2\sqrt{s \mathcal{K}_1} } \bigg\}.
 \label{solution123a}
\end{eqnarray}
To $k_1=0$ in  Eq. (\ref{solution123a}) we  recover the Gaussian solution of Einstein \cite{einstein1905molekularkinetischen}.
To found the complete inversion of Laplace transform to Eqs. (\ref{solution123a}) and (\ref{solution12}) we introduce the Mittag--Leffler function for two parameters \cite{sabatier2007advances}, as follow
\begin{eqnarray}
E_{\alpha,\beta}(z)=\sum_{n=0}^{+\infty} \frac{z^n}{\Gamma[\alpha n + \beta]},
\end{eqnarray}
which obeys the transformations law 
\begin{eqnarray}
\mathcal{L} \{ t^{\beta-1}E_{\alpha,\beta}(-\lambda t^{\alpha}) \} =\frac{s^{\alpha-\beta}}{s^{\alpha}+\lambda}.
\label{mittag}
\end{eqnarray}
 Thereby, we have the inverse solution to Eqs.  (\ref{solution123a})  and (\ref{solution12}) as
\begin{eqnarray}
 \mathcal{W}_1(x,t) &=& \frac{\exp \left[-\frac{(x-l)^2}{4 \mathcal{K}_1 t} \right]}{2\sqrt{\pi t \mathcal{K}_1}} \nonumber \\ 
 & - & \frac{k_1}{2\sqrt{\mathcal{K}_1}}\int_0^t dt' (t-t')^{-\frac{1}{2}}E_{\frac{1}{2},\frac{1}{2}}\left[ -\gamma (t-t')^{\frac{1}{2}} \right]  \frac{\exp \left[- \frac{\left( \left| x\right| +|l| \right)^2}{4 \mathcal{K}_1 t'}  \right]}{2\sqrt{\pi t' \mathcal{K}_1}},
 \label{solution123} \\
 \mathcal{W}_2(x,t) &=& \frac{k_1}{2\sqrt{\mathcal{K}_1}}\int_0^t dt' (t-t')^{-\frac{1}{2}}E_{\frac{1}{2},\frac{1}{2}}\left[ -\gamma (t-t')^{\frac{1}{2}} \right]  \nonumber \\
&\times & \frac{\exp \left[-\frac{\left( \left| x\right| + |l| \sqrt{\mathcal{K}_1^{-1}}\sqrt{\mathcal{K}_2} \right)^2}{4 t' \mathcal{K}_2}  \right] }{2\sqrt{\pi t' \mathcal{K}_2}}.
\end{eqnarray}
 This solution reveals a set of new aspects to distributions of probability to controlled-reaction diffusion problem. In Figs. (\ref{fig2}) we illustrate the temporal evolution of distributions to particles of type $1$ and $2$, that present the relaxation process. Surprisingly, initially, the distributions associated with the walkers of type $1$ and $2$ presents a non-Gaussian characteristic, which does not occur to uniform-linear reaction term  \cite{lenzi2017intermittent}. 
 All the figures that will be presented were generated by numerical integration and special functions (Mittag--Leffler of two parameters, for example) that are contained in the program Wolfram Mathematica 9.

\subsection{\label{sec31} Survival Probability and mean square displacement}

\begin{figure}[h]
\centering
\includegraphics[scale=0.8]{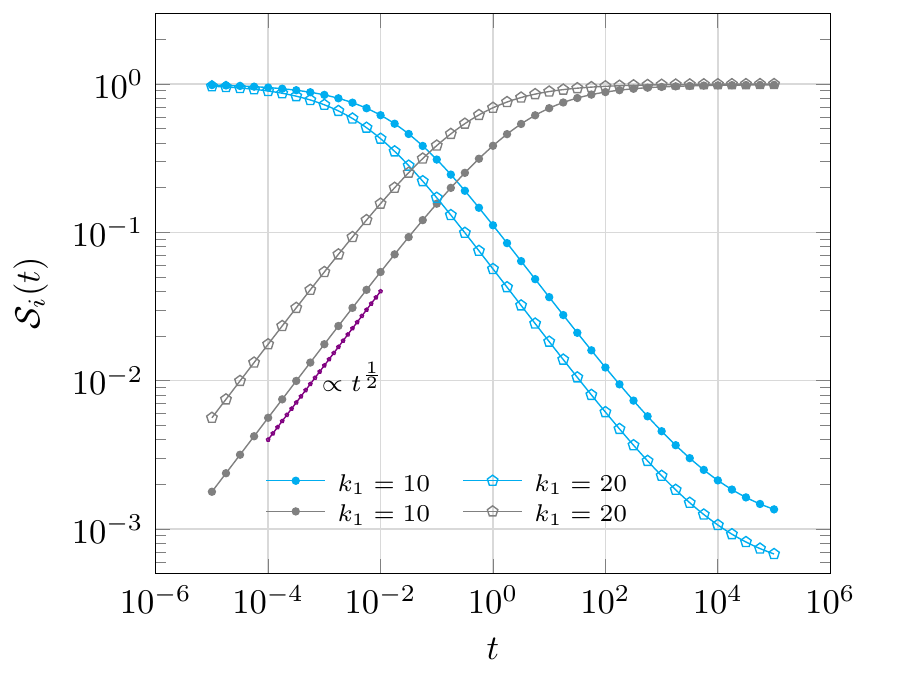}
\caption{\small{The figures present the survival probability for two case, to both walks, for different values of $k_1$ parameter.  We consider $l=0$, $k_1=10$, $k_2=10^{-2}$, $\mathcal{K}_1=1$ and $\mathcal{K}_2=1$ values to particles of type 1 (\textnormal{\textcolor{cyan}{\CIRCLE}}) and type 2 (\textnormal{\textcolor{gray}{\CIRCLE}}).  We consider $l=0$, $k_1=20$, $k_2=10^{-2}$, $\mathcal{K}_1=1$ and $\mathcal{K}_2=1$ values to particles of type 1 ({\normalsize \textcolor{cyan}{$\pentagon$}}) and type 2 ({\normalsize \textcolor{gray}{$\pentagon$}}).
}}
\label{fig4}
\end{figure}

To expand our comprehension about the influence of localised reaction in diffusion context \cite{szabo1984localized,wilemski1973general}. We realise the calculation of survival probability and MSD to particles of type 1 and 2.  
The analytic expression to survival probability  is defined as $
\mathcal{S}_{i}(t)=\int_{-\infty}^{+\infty}dx \mathcal{W}_{i}(x,t)$. This quantity permits us to quantify the fluctuation which occurs on the number of walkers on systems.
 The general expression to survival probability can be written as follow
\begin{eqnarray}
\mathcal{S}_1(t)
&=& 1- \frac{k_1}{2\sqrt{\mathcal{K}_1}} \int_{0}^t dt' t'^{-\frac{1}{2}} E_{\frac{1}{2},\frac{1}{2}}\left[ -\gamma t'^{\frac{1}{2}}\right]\textnormal{erfc}\left(\frac{l}{2 \sqrt{\mathcal{K}_1 (t-t')}}\right),
\end{eqnarray}
 Note that, the survival probability for particles of type 2 can be written as $\mathcal{S}_2(t)=1-\mathcal{S}_1(t)$.   
The Fig. (\ref{fig4}) exemplify the difference and the influence of controlled-diffusion thought survival probability. To analyse how the particles are removed on systems $1$, we consider $k_1 \neq k_2 > 0$, which implies a power--law behaviour to survival probability before the stationary case.

\begin{figure}[h]
\centering
\includegraphics[scale=0.8]{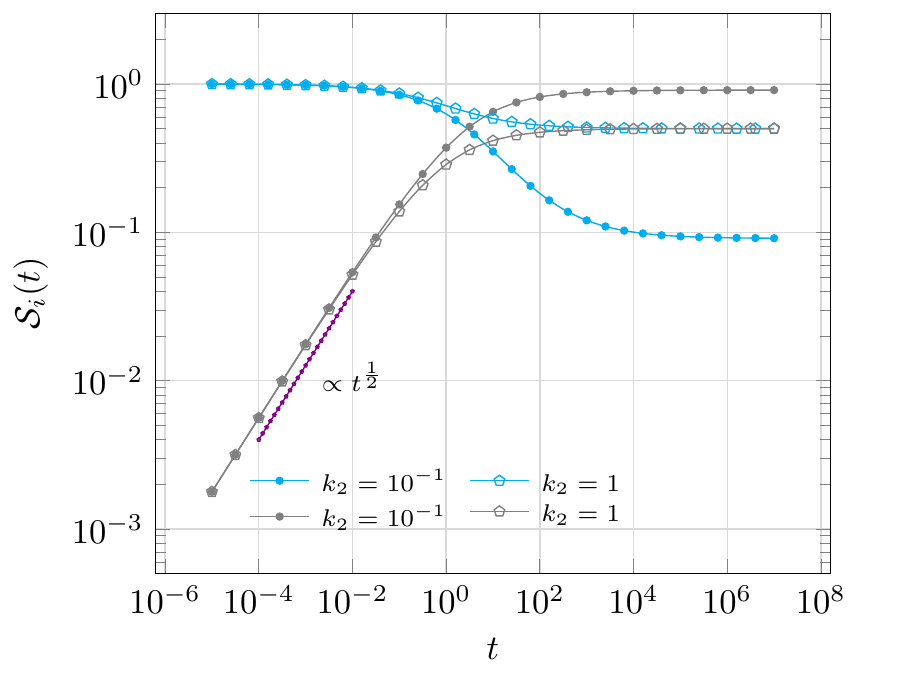}
\caption{\small{The figures present the survival probability for two case, to both walks, for different values of $k_2$ parameter.  We consider $l=0$, $k_1=1$, $k_2=10^{-1}$, $\mathcal{K}_1=1$ and $\mathcal{K}_2=1$ values to particles of type 1 (\textnormal{\textcolor{cyan}{\CIRCLE}}) and type 2 (\textnormal{\textcolor{gray}{\CIRCLE}}).  We consider $l=0$, $k_1=1$, $k_2=1$, $\mathcal{K}_1=1$ and $\mathcal{K}_2=1$ values to particles of type 1 ({\normalsize \textcolor{cyan}{$\pentagon$}}) and type 2 ({\normalsize \textcolor{gray}{$\pentagon$}}).
}}
\label{fig42}
\end{figure}

In the first case, the asymptotic limit of $\mathcal{S}_2$ grows as $\mathcal{S}_2 \propto t^{0.5}$ to short times. And $\mathcal{S}_1=1 - \frac{1}{2\sqrt{\mathcal{K}_1}}\frac{k_1}{\gamma}$ and $\mathcal{S}_2=\frac{1}{2\sqrt{\mathcal{K}_1}}\frac{k_1}{\gamma}$ to long times.  This asymptotic limits reveal that the choice of parameters $k_1$ and $k_2$ change the rate in that particles react on system.

\begin{figure}[h]
\centering
\includegraphics[scale=0.8]{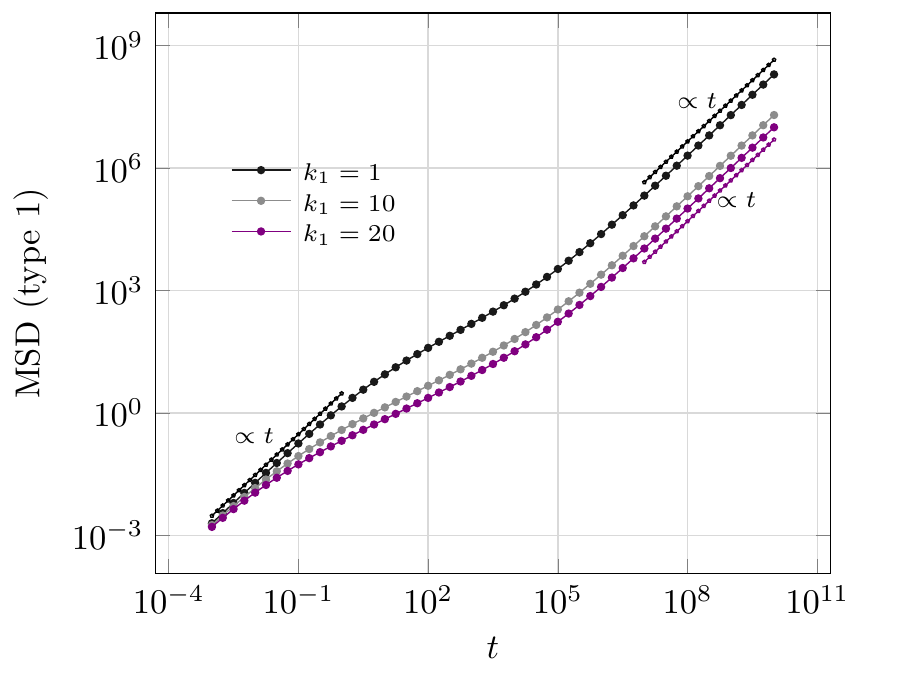} 
\includegraphics[scale=0.8]{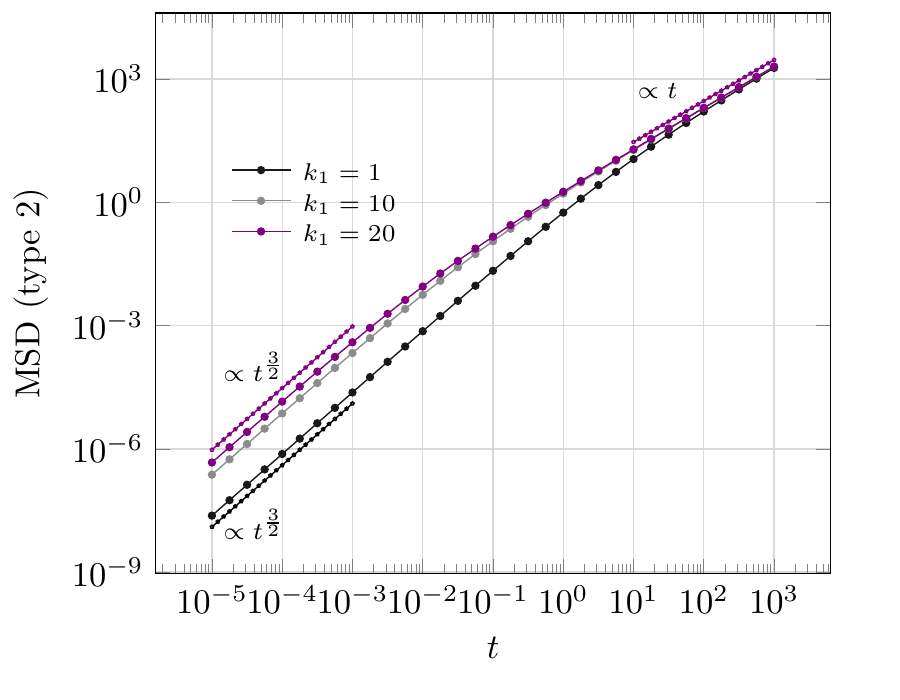} 
\caption{\small{The figure present the MSD to both walkers. Considering $l=0$,   $k_2=10^{-2}$, $\mathcal{K}_1=1$ and  $\mathcal{K}_2=1$.
}}
\label{fig5}
\end{figure}

The MSD behaviour can be calculated by simple integration 
\begin{eqnarray}
\langle (x-\langle x \rangle)^2 \rangle _{i} &=&  \int_{-\infty}^{+\infty} (x-\langle x \rangle _{i})^2 \mathcal{W}_{i}(x,t)dx \nonumber \\
&=& \langle x^2 \rangle _{i} -(2-\mathcal{S}_{i})\langle x \rangle _{i}^2
\label{msds1}
\end{eqnarray}
in which $\langle x \rangle _{i}= \int_{-\infty}^{+\infty} \mathcal{W}_{i}(x,t) x dx $. The distributions (\ref{solution12}) have symmetric form in spacial variable, ergo we have $\langle x \rangle _{2}=0$, on the other hand $\langle x \rangle _{1}=l$. Using the Eq. (\ref{msds1}) and the solutions to first case in Laplace space, we present in Eqs. (\ref{solution11}) and (\ref{solution12}), obtain   
\begin{eqnarray}
\langle x^2 \rangle_1(s) &=&  \frac{2 \mathcal{K}_1}{s^2}\left(1 - \widetilde{\Upsilon}(s) e^{-\sqrt{\frac{s}{\mathcal{K}_1}}|l|} \right) \\
\langle x^2 \rangle_2(s) &=&  \frac{2 \mathcal{K}_2}{s^2}\widetilde{\Upsilon}(s) e^{-\sqrt{\frac{s}{\mathcal{K}_1}}|l|} \label{msds}, 
\end{eqnarray}
using the inverse Laplace transform present in Eq. (\ref{mittag}), we obtain the exact expressions to $\langle x^2 \rangle _{i}$, as follow
\begin{eqnarray}
\langle x^2 \rangle_1(t) &=& 2\mathcal{K}_1t - \frac{k_1}{2\sqrt{\mathcal{K}_1}} \int_0^t dt' 2 \mathcal{K}_1 \tau(t') (t-t')^{-\frac{1}{2}} E_{\frac{1}{2},\frac{1}{2}}\left[ -\gamma (t-t')^{\frac{1}{2}}\right], \\
\langle x^2 \rangle_2(t) &=&  \frac{k_1}{2\sqrt{\mathcal{K}_1}} \int_0^t dt' 2 \mathcal{K}_2 \tau(t') (t-t')^{-\frac{1}{2}} E_{\frac{1}{2},\frac{1}{2}}\left[ -\gamma (t-t')^{\frac{1}{2}}\right] \label{msd},
\end{eqnarray}
in which 
\begin{eqnarray}
\tau(t)=\frac{l \left(\sqrt{\pi } l-2 \sqrt{\mathcal{K}_1 t} e^{-\frac{l^2}{4 \mathcal{K}_1 t}}\right)}{2 \sqrt{\pi } \mathcal{K}_1}+t -\frac{\left(2 \mathcal{K}_1 t+l^2\right) \textnormal{erf}\left(\frac{l}{2 \sqrt{\mathcal{K}_1 t}}\right)}{2 \mathcal{K}_1}
\end{eqnarray}
To exemplify the behaviour of MSD, we consider an reversible condition on Fig. (\ref{fig5}) which illustrate MSD evolution to coupled controlled-diffusion. In this case, the Fig. (\ref{fig5}) that the diffusion of the particle of type 1 have three behaviours, to short time $\langle x^2 \rangle_1(t) \propto t$ (usual--diffusion), the diffusion to times of order $\sim 10^{2}$ the MSD obeys the power--law of the type $\langle x^2 \rangle_1(t) \propto t^{\alpha}$ (sub--diffusion), for long times the diffusive behaviour back to the usual process, i.e. $\langle x^2 \rangle_1(t) \propto t$ (usual--diffusion). To particles of type 2, the MSD behaviour have two regimes, the fractional diffusion to short times $\langle x^2 \rangle_2(t) \propto t^{\frac{3}{2}}$ (super--diffusion), and $\langle x^2 \rangle_2(t) \propto t$ to long times. To found the exact analytical  expression for asymptotic limit for 
long-times, we know that $\lim_{t\rightarrow \infty} \langle (x-\langle x\rangle)_i^2 \rangle   \sim \lim_{t\rightarrow \infty} \langle x^2 \rangle_i$, which implies the follow expression
\begin{eqnarray}
 \lim_{t\rightarrow \infty} \langle x^2 \rangle_i(t) = \mathcal{L}^{-1}\left\{ \lim_{s\rightarrow 0} \mathcal{L}\left\{ \langle x^2 \rangle_i(t)  \right\} \right\}= \frac{\mathcal{K}_i k_j }{\sqrt{\mathcal{K}_j}\left( \frac{k_i}{2\sqrt{\mathcal{K}_i}} + \frac{k_j}{2\sqrt{\mathcal{K}_j}}\right)} t,
 \label{assintotico}
\end{eqnarray}
in which $i \in \{1,2 \}$, $j \in \{1,2\}$, $i \neq j$. If $k_1=0$ in Eq. (\ref{assintotico}) we recover the trivial case that does not occur  the reaction process of particle $1$ for $2$. Therefore, $\langle x^2 \rangle_1(t) \propto 2 \mathcal{K}_1 t$ and $\langle x^2 \rangle_2(t) =0$.

\section{The model with memory effects}
\label{sec4}

The convolution kernel in the diffusion equation is a typical manner of include memory effects \cite{hristov2017derivatives,fractalfract2030020,mafsantos}. The memory kernel can be put in model by two ways, the first one was discussed section \ref{sec2}, the second one was proposed in Ref. \cite{northrup1980stable}, i.e. $\mathcal{R}_{ij} = -\int_0^t dt' k_i(t-t')\rho_i(t')+k_j(t-t') \rho_j(x,t')$. The kernels $k_1(t)$ and $k_2(t)$ represent the memory effects in reaction terms. A important comment here, is necessary emphasise that for  $\mathcal{K}_i(t)=\delta(t)$ we obtain as particular case the exact model of random walks and localised reaction proposed in previous section.  In this scenario, the more general extension of the model proposed by us includes different kernels in reaction point. That is represented by follow equation
\begin{eqnarray}
& & \frac{\partial \ }{\partial t}\mathcal{W}_i(x,t)= \int_{0}^{t}dt' \mathcal{K}_i(t-t')\frac{\partial^2}{\partial x^2} \mathcal{W}_i(x,t') \nonumber  \\
& & - \int_0^t dt'  \delta(x) \left\lbrace k_i(t-t')\mathcal{W}_i(x,t')-k_j(t-t') \mathcal{W}_j(x,t') \right\rbrace,
\label{eq43}
\end{eqnarray}
in which $i \in \{1,2 \}$, $j \in \{1,2\}$, $i \neq j$ and $k_i(t)$ represents the memory kernel associated to the particle of type $i$. This case can include a situation in which the reaction terms go to zero rapidly and decouple the equations, this formulation generalises the previous process to the case where there is a saturation of the quantity of particles that can react in the system. The Laplace--Fourier transform of Eqs. (\ref{eq43}) with conditions (\ref{initialconditions} and \ref{boundariesconditions}) imply in follow equations
\begin{eqnarray}
 \widetilde{\mathcal{W}}_1(k,s)&=& \frac{1}{s+k^2 \widetilde{\mathcal{K}}_1(s)}(e^{-i k l}-\widetilde{k}_1(s) \widetilde{\mathcal{W}}_1(0,s)+\widetilde{k}_2(s) \widetilde{\mathcal{W}}_2(0,s)),  \label{ansatzg}
 \\
  \widetilde{\mathcal{W}}_2(k,s) &=&\frac{1}{s+k^2 \widetilde{\mathcal{K}}_2(s)}(-\widetilde{k}_2(s) \widetilde{\mathcal{W}}_2(0,s)+\widetilde{k}_1(s) \widetilde{\mathcal{W}}_1(0,s)).
  \label{ansatz2g}
\end{eqnarray}
The Solutions in Laplace space has the same structure presented on Eqs. (\ref{ansatz}) and (\ref{ansatz2}), so we have similar set of equations  1 and 2  (\ref{solution11}) and (\ref{solution12}).
 Here will we change the  notations  $k_i \rightarrow \widetilde{k}_i(s)$, $\mathcal{K}_i \rightarrow \widetilde{\mathcal{K}}_i(s)$ and  $\widetilde{\Upsilon}_i(s) \rightarrow \widetilde{\overline{\Upsilon}}_i(s)$ to system with memory, 
we find the following relation to the function $\widetilde{\overline{\Upsilon}}(s)$
\begin{eqnarray}
\widetilde{\overline{\Upsilon}}(s) = \frac{1}{2\sqrt{\widetilde{\mathcal{K}}_1(s)}}\frac{\widetilde{k}_1(s)}{\sqrt{s}+\gamma(s)},
\end{eqnarray}
in which 
\begin{eqnarray}
\gamma(s)= \frac{1}{2}\left(\frac{\widetilde{k}_2(s)}{\sqrt{\widetilde{\mathcal{K}}_2(s)}}+\frac{\widetilde{k}_1(s)}{\sqrt{\widetilde{\mathcal{K}}_1(s)}}\right).
\end{eqnarray}
 Thereby, the solutions in Laplace space can be written as
\begin{eqnarray}
\widetilde{\mathcal{W}}_1(x,s) &=&  \frac{1}{2\sqrt{s\widetilde{\mathcal{K}}_1(s)}}  \exp \left[ -\sqrt{\frac{s}{\widetilde{\mathcal{K}}_1(s)}}|x-l| \right] \nonumber \\ &-& \frac{\widetilde{\overline{\Upsilon}}(s)}{2\sqrt{s\widetilde{\mathcal{K}}_1(s)}}  \exp \left[ -\sqrt{\frac{s}{\widetilde{\mathcal{K}}_1(s)}}(|x|+|l|) \right], \\
\widetilde{\mathcal{W}}_2(x,s) &=&  \frac{\widetilde{\overline{\Upsilon}}(s)}{2\sqrt{s\widetilde{\mathcal{K}}_2(s)}} \exp \left[ -\sqrt{\frac{s}{\widetilde{\mathcal{K}}_2(s)}}|x|- \sqrt{\frac{s}{\widetilde{\mathcal{K}}_1(s)}}|l| \right]. 
\end{eqnarray}

In this part of problem, we obtain a rich class of behaviours. The inverse Laplace transform of these equations are not a trivial problem. To make our analyse we will investigate two cases with memory.

\subsection{First case: $k_i(t)$ memory in rates-reaction}
\label{case1}
\begin{figure}[h]
\centering
\includegraphics[scale=0.8]{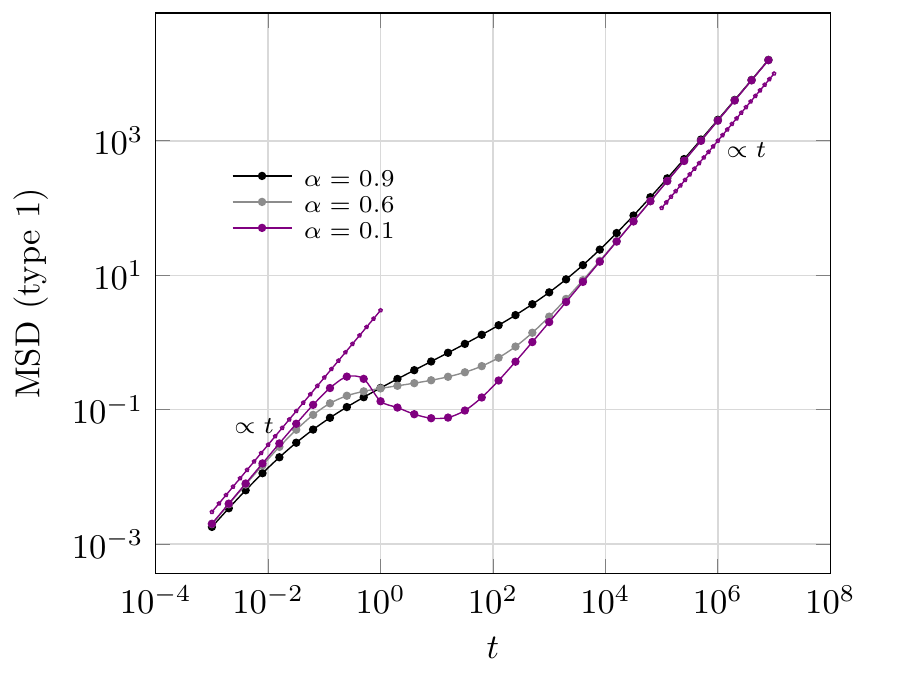} 
\includegraphics[scale=0.8]{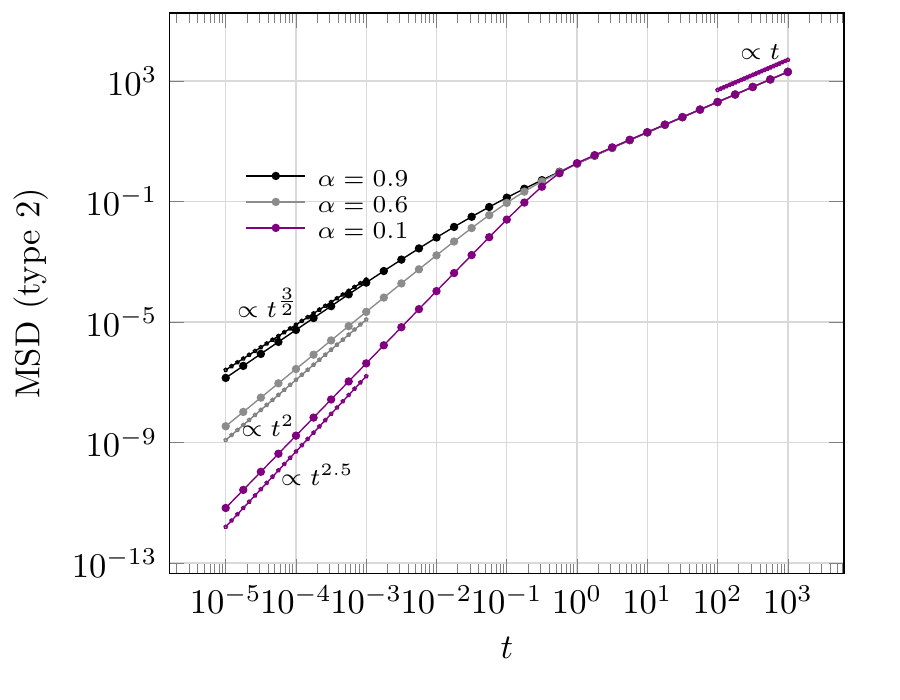}
\caption{\small{The figures present the MSD to both walks for first case with memory, for different values of $\alpha$-index. Considering $c_1=10$, $c_2=10^{-2}$ and $\mathcal{K}_1=\mathcal{K}_2=1$.}}
\label{memory1}
\end{figure}

In this case we consider a constant diffusion coefficient, i.e.  $\mathcal{K}_i=constant$. The choice of kernel investigated here is a power--law functions, \begin{eqnarray}
k_i(t)=\frac{2c_i t^{-\alpha}}{\Gamma(1-\alpha )},
\end{eqnarray}
in which $i \in \{1,2 \}$, $j \in \{1,2\}$, $i \neq j$ and $0<\alpha<1$. However, we denote  $\widetilde{k}_1(s)=2c_1s^{\alpha-1}$ and $\widetilde{k}_2(s)=2c_2s^{\alpha-1}$. We obtain
\begin{eqnarray}
\widetilde{\overline{\Upsilon}}(s) 
&=& \frac{1}{\sqrt{\mathcal{K}_1}}\frac{c_1 }{s^{\frac{3}{2}-\alpha}+\frac{c_2}{\sqrt{\mathcal{K}_2}} +\frac{c_1}{\sqrt{\mathcal{K}_1}} },
\end{eqnarray}
assuming the notation $\beta_1=\frac{c_2}{\sqrt{\mathcal{K}_2}} +\frac{c_1}{\sqrt{\mathcal{K}_1}}$, we obtain the follow expressions 
\begin{eqnarray}
 \mathcal{W}_1(x,t) &=& \frac{\exp \left[-\frac{(x-l)^2}{4 \mathcal{K}_1 t} \right]}{2\sqrt{\pi t \mathcal{K}_1}} \nonumber \\ 
 & - & \frac{c_1}{  \sqrt{\mathcal{K}_1}}\int_0^t dt' (t-t')^{\frac{1}{2}-\alpha}E_{\frac{3}{2}-\alpha,\frac{3}{2}-\alpha}\left[ -\beta_1 (t-t')^{\frac{3}{2}-\alpha} \right] \nonumber \\ 
 &\times & \frac{\exp \left[-\frac{(|x|+|l|)^2}{4 \mathcal{K}_1 t'} \right]}{2\sqrt{\pi t' \mathcal{K}_1}},
 \label{solution123}
\end{eqnarray}
and
\begin{eqnarray}
 \mathcal{W}_2(x,t) &=& \frac{c_1}{\sqrt{\mathcal{K}_1}}\int_0^t dt' (t-t')^{\frac{1}{2}-\alpha}E_{\frac{3}{2}-\alpha,\frac{3}{2}-\alpha}\left[ -\beta_1 (t-t')^{\frac{3}{2}-\alpha} \right] \nonumber \\ &\times & \frac{\exp \left[-\frac{\left( \left| x\right| + |l| \sqrt{\mathcal{K}_1^{-1}}\sqrt{\mathcal{K}_2} \right)^2}{4 t' \mathcal{K}_2}  \right]}{2\sqrt{\pi t' \mathcal{K}_2}}.
\end{eqnarray}
From the point of view of distributions or survival probability for the system with power-law type
memory, the system behaves similarly to the case without memory ($k_i(t)\propto \delta(t)$). However, the dynamics
of the system changes completely, the amount that evidences this difference is MSD, this quantity gives us a more detailed information on how the diffusion process happens to be influenced by the power-law
function to memory kernel present in the terms of reaction. Since the distributions are symmetric to $l=0$, i.e.  $\mathcal{W}_i(x,t)=\mathcal{W}_i(-x,t)$, the MSD assumes the following expressions
\begin{eqnarray}
\langle x^2 \rangle_1 &=& 2\mathcal{K}_1 t - \frac{c_1}{\sqrt{\mathcal{K}_1}}\int_0^t dt' t'^{\frac{1}{2}-\alpha}E_{\frac{3}{2}-\alpha,\frac{3}{2}-\alpha}\left[ -\beta_1 t'^{\frac{3}{2}-\alpha} \right] \times 2\mathcal{K}_1 (t-t'),
\\ 
\langle x^2 \rangle_2 &=& \frac{c_1}{\sqrt{\mathcal{K}_1}}\int_0^t dt' t'^{\frac{1}{2}-\alpha}E_{\frac{3}{2}-\alpha,\frac{3}{2}-\alpha}\left[ -\beta_1 t'^{\frac{3}{2}-\alpha} \right] \times 2\mathcal{K}_2 (t-t'),
\end{eqnarray}
in which $\langle x^2 \rangle_i = \int_{-\infty}^{+\infty} \mathcal{W}_i (x,t) x^2 dx$, to $i=\{1, 2\}$. The figure (\ref{memory1}) exemplifies the effect that a power-law memory has on the local reaction points. In this figure we vary the value of the index $\alpha \in ( 0, 1 ] $ associated with the
memory-kernel, i.e. $k_i(t)\propto \frac{t^{-\alpha}}{\Gamma(1-\alpha )}$, in which for $\alpha=1$ we have the Markovian case (no
memory). The MSD behaviour for the particle of type 1 changes considerably with the presence of memory in the
system, the novelty is that it undergoes a regime in which the MSD decreases before the equations
become decoupled. The particle of type 2 exhibits a hyper diffusive behaviour for short times, and in
general, both substances (type 1 and 2) have a usual behaviour for long times, i.e. $\langle x^2 \rangle \propto t$.

\subsection{Second case: $\mathcal{K}_i(t)$ memory in Laplacian terms}
\label{case2}

In this case we consider $k_i=constant$. The choice of kernel investigated here is a power--law functions,
\begin{eqnarray}
\mathcal{K}_i(t)=\mathcal{D}_{i}\frac{ t^{\mu-1}}{\Gamma(\mu )},
\end{eqnarray}
in which $i \in \{1,2 \}$, $j \in \{1,2\}$, $i \neq j$ and $\frac{1}{2}>\mu>0$. However, we denote  $\widetilde{\mathcal{K}}_1(s)=\mathcal{D}_1 s^{-\mu}$ and $\widetilde{\mathcal{K}}_2(s)=\mathcal{D}_2 s^{-\mu}$. We obtain
\begin{eqnarray}
\widetilde{\overline{\Upsilon}}(s) 
&=& \frac{1}{2\sqrt{\mathcal{D}_1}}\frac{k_1 }{s^{\frac{1-\mu}{2}}+ \beta_2 },
\end{eqnarray}
assuming the notation $\beta_2=\frac{k_2}{2\sqrt{\mathcal{D}_2}} +\frac{k_1}{2\sqrt{\mathcal{D}_1}}$, we obtain the follow expressions  in Laplace space
\begin{eqnarray}
\widetilde{\mathcal{W}}_1(x,s) &=&  \frac{1}{2\sqrt{s^{1-\mu}\mathcal{D}_1}}  \exp \left[ -\sqrt{\frac{s^{1+\mu}}{\mathcal{D}_1}}|x-l|  \right]\nonumber \\ &-& \frac{\widetilde{\overline{\Upsilon}}(s) }{2\sqrt{s^{1-\mu}\mathcal{D}_1}}  \exp \left[ -\sqrt{\frac{s^{1+\mu}}{\mathcal{D}_1}}(|x|+|l|)   \right],\label{s11}
 \\
\widetilde{\mathcal{W}}_2(x,s) &=&  \frac{\widetilde{\overline{\Upsilon}}(s)}{2\sqrt{s^{1-\mu}\mathcal{D}_2}}\exp \left[ -\sqrt{\frac{s^{1+\mu}}{\mathcal{D}_2}}|x|  -\sqrt{\frac{s^{1+\mu}}{\mathcal{D}_1}}|l| \right],
\label{s22}
\end{eqnarray}
using the formula
\begin{eqnarray}
\mathcal{L}^{-1} \left\{ s^{-\sigma} \exp[-z s^{\lambda} ] \right\} = t^{\sigma-1}{\mbox{\Large{H}}}^{1,0}_{1,1}\left[  \frac{|z|}{t^{\lambda}}  \Bigg| ^{\left(\sigma, \lambda \right)}_{(0,1)} \right],
\label{formula}
\end{eqnarray}
in which H is a Fox function. 
Make using of Eq. (\ref{formula}) in Eqs. (\ref{s11}) and (\ref{s22}) we obtain
\begin{eqnarray}
\mathcal{W}_1(x,t) &=&  \frac{1}{2\sqrt{t^{1+\mu}\mathcal{D}_1}} {\mbox{\Large{H}}}^{1,0}_{1,1}\left[  \frac{|x-l|}{\sqrt{\mathcal{D}_1}t^{\frac{1+\mu}{2}}}  \Bigg| ^{\left(\frac{1-\mu}{2}, \frac{1+\mu}{2} \right)}_{(0,1)} \right] \nonumber \\ &-& \int_0 ^t dt' \frac{\overline{\Upsilon}(t-t')}{2\sqrt{t'^{1+\mu}\mathcal{D}_1}} {\mbox{\Large{H}}}^{1,0}_{1,1}\left[  \frac{|x|+|l|}{\sqrt{\mathcal{D}_1}t'^{\frac{1+\mu}{2}}}  \Bigg| ^{\left(\frac{1-\mu}{2}, \frac{1+\mu}{2} \right)}_{(0,1)} \right], \\
\mathcal{W}_2(x,t) &=&  \int_0 ^t dt' \frac{\overline{\Upsilon}(t-t')}{2\sqrt{t'^{1+\mu}\mathcal{D}_2}} {\mbox{\Large{H}}}^{1,0}_{1,1}\left[  \frac{|x|}{\sqrt{\mathcal{D}_2}t'^{\frac{1+\mu}{2}}} + \frac{|l|}{\sqrt{\mathcal{D}_1}t'^{\frac{1+\mu}{2}}}  \Bigg| ^{\left(\frac{1-\mu}{2}, \frac{1+\mu}{2} \right)}_{(0,1)} \right],
\end{eqnarray}
in which 
\begin{eqnarray}
\overline{\Upsilon}_2(t)= \frac{k_1}{2\sqrt{\mathcal{D}_1}}t^{-\frac{1+\mu}{2}} E_{\frac{1-\mu}{2},\frac{1-\mu}{2}}\left[ -\beta_2 t^{\frac{1-\mu}{2}} \right].
\end{eqnarray}

\begin{figure}[h]
\centering
\includegraphics[scale=0.8]{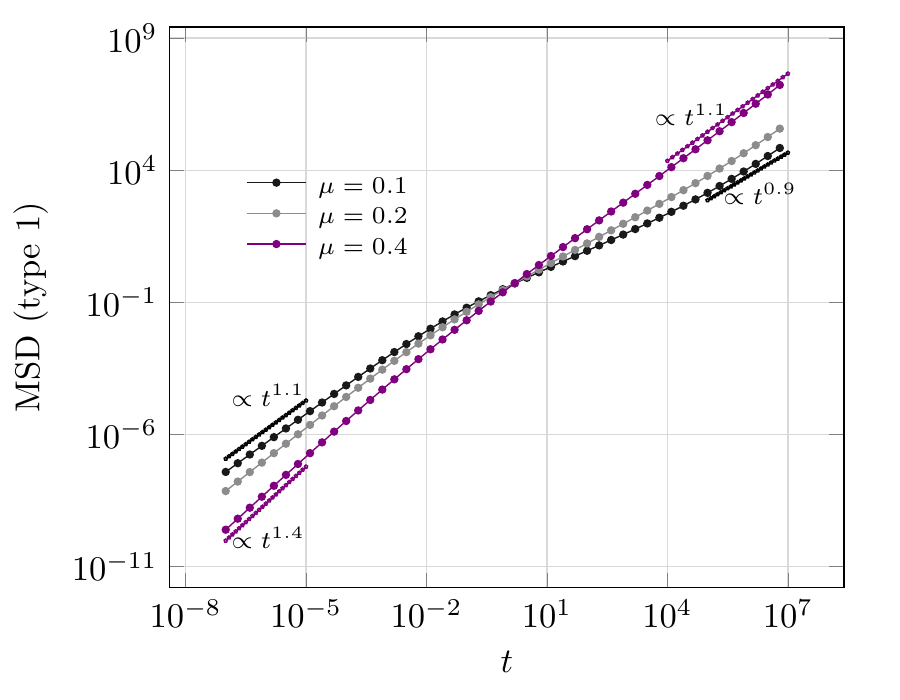} 
\includegraphics[scale=0.8]{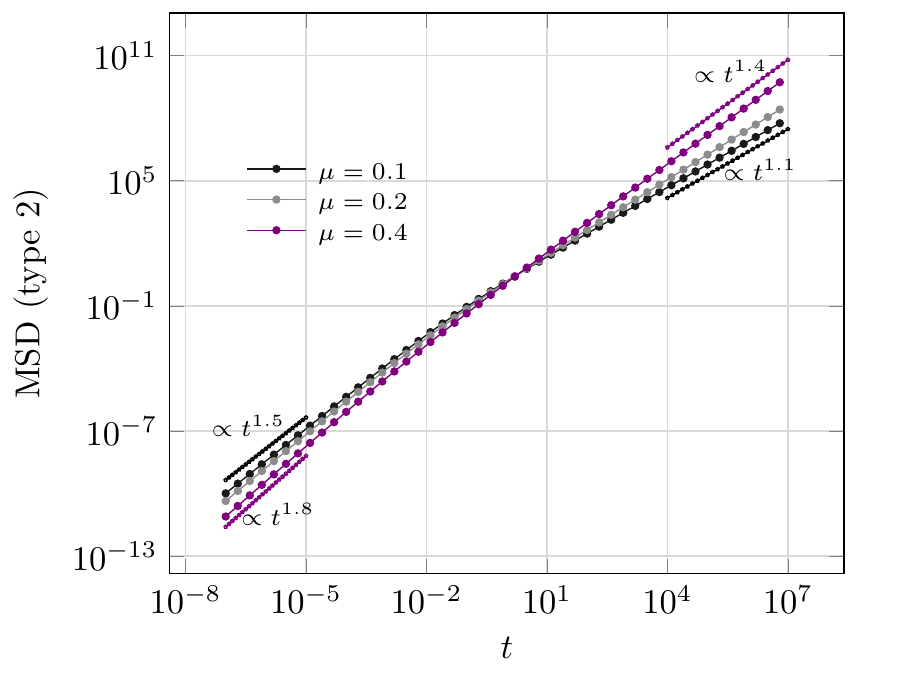}
\caption{\small{The figures present the MSD to both walks for second case with memory, for different values of $\mu$-index. Considering $k_1=10$, $k_2=10^{-2}$ and $\mathcal{D}_1=\mathcal{D}_2=1$.}}
\label{memory2}
\end{figure}

For simplicity we consider $l=0$, the MSD assume the following expressions
\begin{eqnarray}
\langle x^2 \rangle_1 &=& \frac{2 \mathcal{D}_1 t^{1+\mu }}{\Gamma [2+ \mu ]} \nonumber \\ &-& \frac{k_1}{2\sqrt{\mathcal{D}_1}}\int_0^t dt' t'^{-\frac{1+\mu}{2}} E_{\frac{1-\mu}{2},\frac{1-\mu}{2}}\left[ -\beta_2 t'^{\frac{1-\mu}{2}} \right] \times\frac{2 \mathcal{D}_1 (t-t')^{1+\mu }}{\Gamma [2+\mu ]},
\\ 
\langle x^2 \rangle_2 &=& \frac{k_1}{2\sqrt{\mathcal{D}_1}}\int_0^t dt' t'^{-\frac{1+\mu}{2}} E_{\frac{1-\mu}{2},\frac{1-\mu}{2}}\left[ -\beta_2 t'^{\frac{1-\mu}{2}} \right] \times \frac{2 \mathcal{D}_2 (t-t')^{1+\mu }}{\Gamma [2+\mu ]}.
\end{eqnarray}
 The figure (\ref{memory2}) exemplifies the effect that a power-law memory has on the laplacian term. In this figure we vary the value of the index $\mu$ associated with the
memory-kernel, in which for $\mu=0$ we have the Markovian case (no memory). The particle of type 1 have two specific behaviours, to short time we have $\langle x^2 \rangle_1(t) \propto t^{1+\mu}$ and long time $\langle x^2 \rangle_1(t) \propto t^{\frac{1+3\mu}{2}}$ (sub--usual--super diffusion for $k_1\neq 0$). To particles of type 2, the MSD behaviour have two regimes, the fractional diffusion to short times $\langle x^2 \rangle_2(t) \propto t^{\frac{3+\mu}{2}}$   super--diffusion, and $\langle x^2 \rangle_2(t) \propto t^{1+\mu}$ (super--diffusion) to long times. 

Surprisingly, the results presented in the sections \ref{case1} and \ref{case2}  for the MSD to particles of the type $1$ revealed a slow-increasing behaviour (specific time regimes) for some values of the fractional values ($0<\mu<\frac{1}{2}$), this type of regime is not intuitive, since it does not occur in diffusion processes with linear reaction term, i.e., $\mathcal{R}=-k_1 \rho_1 + k_2 \rho_2$, see Ref. \cite{lenzi2017intermittent}. This type of MSD behaviour with \textit{plateau} (three diffusive regimes) occurs in the diffusion of nanoparticles immersed in polymeric media \cite{ge2017nanoparticle,cai2015hopping}, which opens the possibility of investigating disordered structures combined with controlled-diffusion. On the other hand, the MSD for the two systems with localised reactions capture very well the essence of
the problem that is characterised by the presence of plateaus \cite{dos2017anomalous,ahlberg2015many}, thus implying interesting results for the
investigation of localised reaction in the context of model to random walks with traps proposed by Szabol \cite{szabo1984localized}.

The model presented in this work can be used to approach problems of diffusion-controlled reaction \cite{WEISS,szabo1984localized}, coupled reactive Smoluchowski equations \cite{PISERCHIA}, diffusion with time-dependent rate coefficients on reaction \cite{Attila}, and minimal model \cite{flekkoy2017minimal} for two species.

\section{Conclusion}
\label{conclusion}

This paper addressed two areas that are of interest to many researchers. The complexity of controlled-diffusion reaction theory  and the CTRW.
 We considered two species of particles governed by coupled CTRW. The coupling occurs at points representing localised  terms. From the point of view of a single species of particle, these points represent a localised reaction. 

Using mathematical techniques associated with random walks, we exhibited exact solutions to the problem. We presented that solutions are distributions that have non-Gaussian forms, which can be unimodal or not. The analyses performed in the section \ref{sec3} revealed that the system has a rich class of diffusive processes that includes regimes of sub-usual-super diffusion. Moreover, we analysed the probability of survival to both species of particles. We demonstrated that before the moment in which a total balance occurs between the species, for long times the probability of survival follows a very specific type of power--law. 

Finally, we generalised the model in order to include memory terms at localised reaction points and in laplacian term. We obtained the exact solutions and MSD when memory kernels are power--law functions. We have shown that the coupling of random walkers differently affect the behaviour of MSD. In particular, we obtained a counterintuitive behaviour given by quasi-constant MSD regimes (plateau regimes) which is a well-known behaviour for the diffusion of nanoparticles in polymeric structures. 
 
 The methods and the techniques applied in this work allow a new approach to the problems associated with precipitation of particles on a surface with irregularities in which irregularities act as reaction-traps. In addition, this work opens possibilities to investigate the coupled controlled-diffusion reaction in  the context of fractional derivatives  or memory function associated with special functions \cite{agarwal2018solutions,garra2018prabhakar}.
 

\section*{Acknowledgements} 

We thank the Brazilian agency CNPq.

\section*{References}

\bibliographystyle{iopart-num}

\end{document}